\documentclass[a4paper,12pt]{article}
\usepackage{amssymb}
\usepackage{amsmath}
\usepackage{ascmac}
\usepackage{comment}
\usepackage{wrapfig}
\usepackage{ulem}
\newtheorem{thm}{Theorem}[subsection]
\newtheorem{prop}[thm]{Proposition}
\newtheorem{lem}[thm]{Lemma}

\newtheorem{rem}[thm]{Remark}

\def\C{{\mathbb C}}

\def\Z{{\mathbb Z}}
\def\N{{\mathbb N}}
\def\dfrac#1#2{{\displaystyle\frac{#1}{#2}}}

\def\o#1{{\overline{#1}}}
\def\u#1{{\underline{#1}}}
\def\ds{\displaystyle}
\def\prf{\noindent{\bf Proof.} }
\def\qed{\hfill$\blacksquare$}
\def\sq{\hfill$\square$}

\makeatletter
\@addtoreset{equation}{section}

\makeatother

\begin{document}

\begin{center}
{\Large {\bf Variations of $q$-Garnier system}} 

\vspace{5mm}
{\large By}

\vspace{5mm}
{\large Hidehito Nagao$^1$ and Yasuhiko Yamada$^2$}

(Akashi College$^1$ and Kobe University$^2$, Japan)\\
\today
\end{center}

\noindent
{\bf Abstract.}
We study several variants of q-Garnier system corresponding to various  
directions of discrete time evolutions.
We also investigate a relation between the $q$-Garnier system and Suzuki's higher order $q$-Painlev\'e system by using a duality of the $q$-KP system.

\vspace{5mm}



\renewcommand\baselinestretch{1.2}
\section{Introduction}\label{sect:intro}

The $q$-Garnier system is a multivariable extension of the $q$-Painlev\'e ${\rm VI}$ equation \cite{JS96}. It was first formulated in \cite{Sakai05-1} and has been studied in \cite{NY16, OR16-1, Sakai05-2} from differing points of view.\footnote{Other discrete analogues such as  elliptic and additive Garnier systems have been studied in \cite{ND16,OR16-2, Yamada17} and \cite{DST13, DT14}.}
For continuous/discrete Painlev\'e and Garnier systems in general, one can consider various directions of 
deformations (evolutions).
In this paper, we consider such variations for  $q$-Garnier system.\footnote{For differential Garnier system, the ``{\it continuous time evolutions}" and  the ``{\it Schlesinger transformations}" are usually considered as different kind of equations. However, in difference case, we call both of them ``{\it Garnier system}", since we can treat them in similar manner.} 
Our motivations and results are as follows.
\begin{itemize}
\item In general, we need to choose dependent variables appropriately according to their directions 
in order to make  evolution equations in tractable form. 
We will derive simple equations  for fundamental directions of $q$-Garnier systems by using the contiguity type scalar Lax pairs.


\item  In \cite{NY16} a connection between Suzuki's higher order $q$-Painlev\'e system \cite{Suzuki15, Suzuki17} and $q$-Garnier system is suggested  by comparing their special solutions.
We clarify the connection by using a duality of the $q$-KP system \cite{KNY02-1, KNY02-2}.

\end{itemize}

This paper is organized as follows. In Section \ref{sec:Gar}, based on our previous work \cite{NY16}, we rewrite Sakai's $2 \times 2$ matrix Lax form into a scalar Lax form. In Section \ref{sec:Vn}, we exemplify the scalar Lax pairs and the evolution equations for the various $q$-Garnier systems, by choosing the proper dependent variables. In Section \ref{sec:Rn}, as an application of the
results in Section \ref{sec:Vn}, we give two reductions from case $N=3$ of the various $q$-Garnier systems to $q$-Painlev\'e  systems of type $E_7^{(1)}$. In Section \ref{sec:KP}, extending the duality  for reduced $q$-KP system \cite{KNY02-1, KNY02-2}, we show the equivalence between Suzuki's system and a certain variation of $q$-Garnier system. 

\section{$q$-Garnier system}\label{sec:Gar}　
In this section we recall Sakai's $2 \times 2$ matrix Lax from \cite{Sakai05-1} for the $q$-Garnier system and its corresponding scalar Lax form \cite{NY16}. 
\subsection{Sakai's matrix Lax form}\label{subsec:Gar_mat}
In \cite{Sakai05-1}, the $q$-Garnier system has been formulated by H.Sakai. For an unknown function $Y(z)={}^t\left[y_1(z), y_2(z)\right]$, the $2 \times 2$ matrix Lax form for the $q$-Garnier system is given by
\begin{equation}\label{eq:Gar_mat_L}
Y(qz)={\mathcal A}(z)Y(z), 
\end{equation}
\begin{equation}\label{eq:Gar_mat_D}
\o{Y}(z)={\mathcal B}(z)Y(z).
\end{equation}
Here $q$ is a (complex) parameter, and ${\mathcal A}$ and ${\mathcal B}$ are $2 \times 2$ matrices depending on $z$ and other parameters. The matrix ${\mathcal A}$ is a polynomial in $z$ (see properties (\ref{eq:Gar_mat_A}) below), on the other hand the matrix ${\mathcal B}$ is rational in $z$. We denote the time evolution by $T(*)=\o{*}$. The equation (\ref{eq:Gar_mat_L}) gives the linear $q$-difference equation and the equation (\ref{eq:Gar_mat_D}) gives its deformation equation\footnote{The matrix ${\mathcal B}$ can be chosen according to various deformation directions.}.
The compatibility condition for the matrix systems (\ref{eq:Gar_mat_L}) and (\ref{eq:Gar_mat_D}) reads 
\begin{equation}\label{eq:Gar_mat_com}
\o{\mathcal A}(z){\mathcal B}(z)={\mathcal B}(qz){\mathcal A}(z), 
\end{equation}
which gives the $q$-Garnier system. Let $\alpha_1$, \ldots $\alpha_{2N+2}$, $\kappa_1$, $\kappa_2$, $\theta_1$ and $\theta_2$ be (complex) parameters. 
Then the matrix ${\mathcal A}$ is characterized by the following properties :
\begin{equation}\label{eq:Gar_mat_A}
\begin{array}l
\ds {\rm (i)} \quad {\mathcal A}(z)=A_0+A_1z+\cdots +A_{N+1}z^{N+1}, \\[5mm]
\ds {\rm(ii)} \quad A_{N+1}=\left[\begin{array}{cc}\kappa_1&0\\ 0&\kappa_2\end{array}\right], \quad \mbox{$A_0$ has eigenvalues $\theta_1$ and $\theta_2$}
,\\[5mm]
\ds {\rm(iii)} \quad |{\mathcal A}(z)|=\kappa_1\kappa_2\prod_{i=1}^{2N+2}(z-\alpha_i),\quad (\kappa_1\kappa_2\prod_{i=1}^{2N+2}\alpha_i=\theta_1\theta_2).
\end{array}
\end{equation} 

The matrix ${\mathcal A}$ satisfying these properties (i)$-$(iii) has $2N+1$ arbitrary parameters except parameters $\{\alpha_i, \kappa_i, \theta_i\}$. The $2N$ parameters of them can be interpreted as variables for the $q$-Garnier system and the remaining parameter is a ``{\it gauge}" freedom. The deformation direction in \cite{Sakai05-1} is given by $\o{\alpha}_i=q\alpha_i$ $\o{\alpha}_j=\alpha_j$, $\o{\kappa}_i=\kappa_i$ and $\o{\theta}_i=q\theta_i$ $(i=1,2. \quad j=3,\ldots, 2N+2)$. 

We must choose proper coordinates of variables according to the deformation directions, in order to obtain simple forms of the $q$-Garnier systems. This content will be shown in Section \ref{sec:Vn}.

\subsection{Scalar Lax form}\label{subsec:Gar_Sca}
We recall the scalar Lax form and derive the simple form of the time evolution equation from its compatibility condition.   

\begin{lem}\label{lem:Gar_Sca_L}
The matrix system (\ref{eq:Gar_mat_L}) and (\ref{eq:Gar_mat_D}) can be rewritten into the following linear $q$-difference equations: 
\begin{equation}\label{eq:Gar_Sca_L}
\begin{array}l
\ds L_1: F(\frac{z}{q})y(qz)-\left\{F(\frac{z}{q}){\mathcal A_{11}}(z)+F(z){\mathcal A_{22}}(\frac{z}{q})\right\}y(z)+F(z)|{\mathcal A}(\frac{z}{q})|y(\frac{z}{q})=0,\\[3mm]
L_2: F(z)\o{y}(z)-H(z)y(qz)+G(z)y(z)=0,\\[3mm]
L_3: \o{F}(z)|{\mathcal B}(z)|y(qz)-H(qz)|{\mathcal A}(z)|\o{y}(z)+G(z)\o{y}(qz)=0.
\end{array}
\end{equation}
Here 
\begin{equation}\label{eq:Gar_Sca_FGH}
y(z)=y_1(z), \hspace{1mm} F(z)={\mathcal A_{12}}(z), \hspace{1mm} G(z)={\mathcal A}_{11}(z){\mathcal B}_{12}(z)-{\mathcal A}_{12}(z){\mathcal B}_{11}(z), \hspace{1mm} H(z)={\mathcal B}_{12}(z).
\end{equation}
\sq  
\end{lem}

\prf
We write the matrix system (\ref{eq:Gar_mat_L}) as 
\begin{equation}\label{eq:Gar_Sca_A}
  \begin{cases}
    y_1(qz)={\mathcal A_{11}}(z)y_1(z)+{\mathcal A_{12}}(z)y_2(z), & \\
    y_2(qz)={\mathcal A_{21}}(z)y_1(z)+{\mathcal A_{22}}(z)y_2(z). &
  \end{cases}
\end{equation}
Eliminating the component $y_2(z)$ from the equation (\ref{eq:Gar_Sca_A}), we obtain the  linear equation $L_1$ (\ref{eq:Gar_Sca_L}).
The matrix system (\ref{eq:Gar_mat_D}) is also rewritten as
\begin{equation}\label{eq:Gar_Sca_B}
  \begin{cases}
    \o{y}_1(z)={\mathcal B_{11}}(z)y_1(z)+{\mathcal B_{12}}(z)y_2(z), & \\
    \o{y}_2(z)={\mathcal B_{21}}(z)y_1(z)+{\mathcal B_{22}}(z)y_2(z). &
  \end{cases}
\end{equation}
Eliminating the component $y_2(z)$ from the first equations in (\ref{eq:Gar_Sca_A}) and (\ref{eq:Gar_Sca_B}), we obtain the deformation equation $L_2$ (\ref{eq:Gar_Sca_L}). 

On the other hand, rewriting (\ref{eq:Gar_mat_L}) and (\ref{eq:Gar_mat_D}) into 
\begin{equation}\label{eq:Gar_Sca_LD}
\o{Y}(qz)=\o{\mathcal A}(z)\o{Y}(z), \quad \o{Y}(qz)={\mathcal B}(qz)Y(qz),
\end{equation}
and eliminating $y_2(qz)$, $\o{y}_2(z)$ and $\o{y}_2(qz)$ from the system (\ref{eq:Gar_Sca_LD}), then we have 
\begin{equation}\label{eq:Gar_Sca_LL3}
\o{\mathcal A}_{12}(z)|{\mathcal B}(qz)|y(qz)-H(qz)|\o{\mathcal A}(z)|\o{y}(x)-\left\{{\mathcal A}_{12}(z){\mathcal B}_{22}(qz)-\o{\mathcal A}_{22}(z){\mathcal B}_{12}(z)\right\}\o{y}(qz)=0.
\end{equation}
Moreover applying the compatibility condition (\ref{eq:Gar_mat_com}) to (\ref{eq:Gar_Sca_LL3}), we we can derive the deformation equation $L_3$ (\ref{eq:Gar_Sca_L}). 
\qed

We remark that unknown variables of $q$-Garnier systems appear as the coefficients in $z$ of the polynomials $F$, $G$ and $H$.
The polynomial $F$ is of degree $N$ in $z$ and the $N$ points such that $F(z)=0$ are not singularities but``{\it apparent singularities}" (i.e. the solutions are regular there). This means that the two relations 
\begin{equation}\label{eq:Gar_Sca_apparent}
\begin{array}l
\ds\frac{y(qz)}{y(z)}={\mathcal A}_{11}(z),\quad \mbox{for}\quad F(z)=0, \\[5mm]
\ds\frac{y(z)}{y(\frac{z}{q})}=\frac{|{\mathcal A}(\frac{z}{q})|}{{\mathcal A}_{22}(\frac{z}{q})},\quad \mbox{for}\quad F(\frac{z}{q})=0,
\end{array}
\end{equation}
derived 
from the $L_1$ equation (\ref{eq:Gar_Sca_L}) are consistent with each other.

A pair of the deformation equations $\{L_2,L_3\}$ is equivalent to the scalar Lax pair $\{L_1,L_2\}$, so we call the pair $\{L_2,L_3\}$ the ``{\it scalar Lax pair (of contiguous relation type)}". We conveniently deal with the Lax pair $\{L_2,L_3\}$ \cite{NTY13}, since the explicit form of the $L_1$ equation is rather complicated. 

\begin{lem}\label{lem:Gar_Sca_ev}
The compatibility of the scalar Lax pair $\{L_2, L_3\}$ (\ref{eq:Gar_Sca_L}) gives the following relation:
\begin{equation}\label{eq:Gar_Sca_ev}
\begin{array}l
G(z)\u{G}(z)=H(z)\u{H}(qz)|\u{\mathcal A}(z)|, \quad \mbox{for}\quad F(z)=0,\\[3mm]
|{\mathcal B}(z)|F(z)\o{F}(z)=H(z)H(qz)|{\mathcal A}(z)|, \quad \mbox{for}\quad G(z)=0.
\end{array}
\end{equation}
\sq  
\end{lem}

\prf
Under the condition $F(z)=0$, eliminating $y(z)$ and $y(qz)$ from $L_2(z)=\u{L}_3(qz)=0$, we obtain the first relation of (\ref{eq:Gar_Sca_ev}). Similarly, for $G(z)=0$, eliminating $y(qx)$ and $\o{y}(x)$ from $L_2(z)=L_3(qz)=0$, we have the second relation of (\ref{eq:Gar_Sca_ev}).
\qed

The relation (\ref{eq:Gar_Sca_ev}) may be insufficient for the compatibility of the Lax pair $\{L_2, L_3\}$.  
Additional conditions for the sufficiency will be considered case by case in Section \ref{sec:Vn}. 

Though we have derived the scalar Lax pair from matrix Lax form, the scalar Lax pair $\{L_2, L_3\}$ can be also obtained from a certain method using Pad\'e interpolation (see for example \cite{NY16}). 
In the following sections, we will discuss based on the Lax pair obtained by Pad\'e method\footnote{\label{footnote:Pade}For the proof of the correspondence between (\ref{eq:Gar_Sca_L}) and (\ref{eq:Vn_Da_T3_L1}) independent of Pad\'e method, see Appendix \ref{subsec:Lax_Corres} as an example}. 

\section{Variations of the $q$-Garnier system}\label{sec:Vn}

In this section, we consider Lax pairs and evolution equations for several directions of $q$-Garnier systems. 

\subsection{Notation}\label{subsec:Vn_NR}

Fix a positive integer $N$ and a parameter $q$.
Let 
$a_1, \ldots, a_{N+1}$, $b_1, \ldots, b_{N+1}$, $c_1, c_2, d_1, d_2$ be
parameters with a constraint $\prod_{i=1}^{N+1}\frac{a_i}{b_i}=q\prod_{i=1}^{2}\frac{c_i}{d_i}$ and $T_a: a \mapsto q a$ be the $q$-shift operator of parameter $a$. 

We consider the following four directions $T_1, \ldots, T_4$ defined by
\begin{equation}\label{eq:Vn_NR_T}
\begin{array}l
T_1=T_{a_1}^{-1}T_{b_1}^{-1}, \quad T_2=T_{a_1}^{-1}T_{a_2}^{-1}T_{b_1}^{-1}T_{b_2}^{-1}, \quad T_3=T_{c_1}T_{d_1}, \quad  T_4=T_{a_{N+1}}^{-1}T_{c_1}^{-1},
\end{array}
\end{equation}
and the corresponding shifts are denoted as $\o{X}:=T_i(X)$ and $\u{X}:=T_i^{-1}(X)$. The operators $T_i$ play the role of time evolutions of the $q$-Garnier system.

The directions $T_1$, $T_3$ and $T_4$ (together with related ones obtained by obvious symmetry: $T_{a_i}^{-1}T_{b_j}^{-1}$, $T_{c_i}T_{d_j}$, $T_{a_i}^{-1}T_{c_j}^{-1}$) are fundamental ones, and others (e.g. $T_2$) are given by compositions of them
(and their inverses).

In this subsection, we use the following notations: 
\begin{equation}\label{eq:Vn_NR_AB}
\begin{array}l
\ds A(z)=\prod_{i=1}^{N+1}(z-a_i), \quad B(z)=\prod_{i=1}^{N+1}(z-b_i), \quad A_i(z)=\frac{A(z)}{z-a_i}, \quad B_i(z)=\frac{B(z)}{z-b_i}, \\[3mm]
\ds A_{i,j}(z)=\frac{A(z)}{(z-a_i)(z-a_j)}, \quad B_{i,j}(z)=\frac{B(z)}{(z-b_i)(z-b_j)}, \quad F(z)=\sum_{i=0}^N f_i z^i, \\[3mm]
\end{array}
\end{equation}
where $f_0, \ldots, f_N$ are variables depending on parameters $a_i, b_i, c_i$ and $d_i$.\\

The following data will be described by the polynomials $F(z)$, $G(z)$ and $H(z)$. The polynomial $F$ is common for all cases as given above, while one should take the polynomials $G$ and $H$ differently case by case as given below. 

\subsection{Results for each direction}\label{subsec:Vn_Da}
In the following subsubsections, we show the three items\footnote{The data for each items is obtained  by considering a same Pad\'e problem with various deformations (see Appendix \ref{sec:Pade}).}: (a) Scalar Lax pair, (b) Time evolution equation, (c) The $L_1$ equation. 

In item (b), we give the time evolution equation as necessary and sufficient condition for compatibility of the scalar Lax pair (For the proof of the sufficiency, see Appendix \ref{subsec:Lax_Suf}). In item (c), the $L_1$ equations
in each subsections
are expressed in different forms, however, we will show that they are equivalent with each other
(Theorem \ref{thm:Vn_NR_Gtrans}). 

\subsubsection{
Direction $T_1=T_{a_1}^{-1}T_{b_1}^{-1}$}\label{subsubsec:Vn_Da_T1} 

This case is considered in \cite{NY16} and the direction $T_1$ is (an inverse of) the original direction in \cite{Sakai05-1}.

\noindent
(a) Scalar Lax pair
\begin{equation}\label{eq:Vn_Da_T1_L2L3} 
\begin{array}l
\ds L_2(z)=F(z)\o{y}(z)-A_1(z)y(qz)+(z-b_1)G(z)y(z),\\[5mm]
\ds L_3(z)=\o{F}(\frac{z}{q})y(z)+(z-a_1)G(\frac{z}{q})\o{y}(z)-qc_1c_2B_1(\frac{z}{q})\o{y}(\frac{z}{q}),
\end{array}
\end{equation}
where $A_1(z)$, $B_1(z)$, $F(z)$ are as in (\ref{eq:Vn_NR_AB}) and $G(z)=\sum_{i=0}^{N-1} g_i z^{i}$ .

\noindent
(b) Time evolution equation
\begin{equation}\label{eq:Vn_Da_T1_ev}
\begin{array}l
G(z)\u{G}(z)=c_1c_2\dfrac{A_1(z)B_1(z)}{(z-a_1)(z-b_1)},
\quad {\rm for} \quad F(z)=0,\\[3mm]
F(z)\o{F}(z)=qc_1c_2A_1(z)B_1(z),
\quad {\rm for} \quad G(z)=0,\\[3mm]
\ds f_N\o{f}_N=q (g_{N-1}-c_1)(g_{N-1}-c_2),\\[3mm]
\ds f_0\o{f}_0=a_1b_1\Big(g_0-\frac{d_1}{a_1b_1}A(0)\Big)\Big(g_0-\frac{d_2}{a_1b_1}A(0)\Big),
\end{array}
\end{equation} 
where $2N$ variables $\frac{f_1}{f_0}$,$\ldots$, $\frac{f_N}{f_0}$, $g_0$,$\ldots$, $g_{N-1}$ are the dependent variables.

\noindent
(c) The $L_1$ equation
\begin{equation}\label{eq:Vn_Da_T1_L1}
\begin{array}{l}
\ds L_1(z)=A(z)F(\frac{z}{q})y(q z)+qc_1c_2B(\frac{z}{q})F(z)y(\frac{z}{q})\\[5mm]
 \ds \phantom{\ds L_1(z)} -\Big\{(z-a_1)(z-b_1)F(\frac{z}{q})G(z)+\dfrac{F(z)}{G(\frac{z}{q})}V(\frac{z}{q})\Big\}y(z),
\end{array}
\end{equation}
where $V(z)=qc_1c_2A_1(z)B_1(z)-F(z)\o{F}(z)$.

\subsubsection{
Direction $T_2=T_{a_1}^{-1}T_{a_2}^{-1}T_{b_1}^{-1}T_{b_2}^{-1}$
}\label{subsubsec:Vn_Da_T2}

\noindent
(a) Scalar Lax pair
\begin{equation}\label{eq:Vn_Da_T2_L2L3} 
\begin{array}l
\ds L_2(z)=F(z)\o{y}(z)-A_{1,2}(z)(1+hz)y(qz)+(z-b_1)(z-b_2)G(z)y(z),\\[5mm]
\ds L_3(z)=\o{F}(\frac{z}{q})y(z)+(z-a_1)(z-a_2)G(\frac{z}{q})\o{y}(z)-qc_1c_2B_{1,2}(\frac{z}{q})(1+hz)\o{y}(\frac{z}{q}),
\end{array}
\end{equation}
where $A_{1,2}(z)$, $B_{1,2}(z)$, $F(z)$ are as in (\ref{eq:Vn_NR_AB}) and $G(z)=\sum_{i=0}^{N-2} g_i z^{i}$.

\noindent
(b) Time evolution equation
\begin{equation}\label{eq:Vn_Da_T2_ev}
\begin{array}l
\ds\frac{G(z)\u{G}(z)}{(1+hz)(1+q\u{h}z)}=\frac{c_1c_2A_{1,2}(z)B_{1,2}(z)}{q(z-a_1)(z-a_2)(z-b_1)(z-b_2)},
\quad {\rm for} \quad F(z)=0,\\[5mm]
F(z)\o{F}(z)=qc_1c_2A_{1,2}(z)B_{1,2}(z)(1+hz)(1+qhz),
\quad {\rm for} \quad G(z)=0,\\[5mm]
F(z)\o{F}(\frac{z}{q})=(z-a_1)(z-a_2)(z-b_1)(z-b_2)G(z)G(\frac{z}{q}),
\quad {\rm for} \quad 1+hz=0,\\[3mm]
\ds f_N\o{f}_N=q^2 (g_{N-2}-c_1h)(g_{N-2}-c_2h),\\[3mm]
\ds f_0\o{f}_0=a_1a_2b_1b_1\Big(g_0-\frac{d_1A(0)}{a_1a_2b_1b_2}\Big)\Big(g_0-\frac{d_2A(0)}{a_1a_2b_1b_2}\Big),
\end{array}
\end{equation}
where $2N$ variables $\frac{f_1}{f_0}$,$\ldots$, $\frac{f_N}{f_0}$, $g_0$,$\ldots$, $g_{N-2}$ and $h$ are the dependent variables.

\noindent
(c) The $L_1$ equation
\begin{equation}\label{eq:Vn_Da_T2_L1}
\begin{array}{l}
\ds L_1(z)=A(z)F(\frac{z}{q})y(q z)+qc_1c_2B(\frac{z}{q})F(z)y(\frac{z}{q})\\[5mm]
\ds \phantom{L_1(z)} -\frac{1}{1+hz}\Big\{(z-a_1)(z-b_1)(z-a_2)(z-b_2)F(\frac{z}{q})G(z)+\dfrac{F(z)}{G(\frac{z}{q})}V(\frac{z}{q})\Big\}y(z),
\end{array}
\end{equation}
where $V(z)=qc_1c_2(1+hz)(1+qhz)A_{1,2}(z)B_{1,2}(z)-F(z)\o{F}(z)$. 

\subsubsection{
Direction $T_3=T_{c_1}T_{d_1}$}\label{subsubsec:Vn_Da_T3}
The deformation direction $T_3$ is the same as that of Suzuki's system \cite{Suzuki15} (see \S\ref{subsec:SD_HOP}). 

\noindent
(a) Scalar Lax pair
\begin{equation}\label{eq:Vn_Da_T3_L2L3} 
\begin{array}l
\ds L_2(z)=F(z)\o{y}(z)-A(z)y(qz)+G(z)y(z),\\
\ds L_3(z)=\frac{z}{q}\o{F}(\frac{z}{q})y(z)+G(\frac{z}{q})\o{y}(x)-qc_1c_2B(\frac{z}{q})\o{y}(\frac{z}{q}),
\end{array}
\end{equation}
where $A(z)$, $B(z)$, $F(z)$ are as in (\ref{eq:Vn_NR_AB}) and $G(z)=\sum_{i=0}^{N+1} g_i z^{i}$, $g_{N+1}=\frac{c_2g_0}{d_1A(0)}$.

\noindent
(b) Time evolution equation
\begin{equation}\label{eq:Vn_Da_T3_ev}
\begin{array}l
\ds G(z)\u{G}(z)=c_1c_2A(z)B(z), \quad {\rm for} \quad F(z)=0,\\[5mm]
\ds zF(z)\o{F}(z)=qc_1c_2A(z)B(z), \quad {\rm for} \quad G(z)=0,\\[5mm]
\end{array}
\end{equation}
where $2N$ variables $\frac{f_1}{f_0}$,$\ldots$, $\frac{f_N}{f_0}$, $\frac{g_1}{g_0}$,$\ldots$, $\frac{g_N}{g_0}$ are the dependent variables.

\noindent
(c) The $L_1$ equation
\begin{equation}\label{eq:Vn_Da_T3_L1}
\begin{array}{l}
\ds L_1(z)=A(z)F(\frac{z}{q})y(q z)+qc_1c_2B(\frac{z}{q})F(z)y(\frac{z}{q})\\[5mm]
\ds \phantom{L_1(z)} -\Big\{F(\frac{z}{q})G(z)+\dfrac{F(z)}{G(\frac{z}{q})}V(\frac{z}{q})\Big\}y(z),
\end{array}
\end{equation}
where $V(z)=qc_1c_2A(z)B(z)-zF(z)\o{F}(z)$. 

\subsubsection{
Direction $T_4=T_{a_{N+1}}^{-1}T_{c_1}^{-1}$}\label{subsubsec:Vn_Da_T4}

\noindent
(a) Scalar Lax pair
\begin{equation}\label{eq:Vn_Da_T4_L2L3} 
\begin{array}l
\ds L_2(z)=F(z)\o{y}(z)-A_{N+1}(z)y(qz)+G(z)y(z),\\
\ds L_3(z)=\o{F}(\frac{z}{q})y(z)+(z-a_{N+1})G(\frac{z}{q})\o{y}(x)-qc_1c_2B(\frac{z}{q})\o{y}(\frac{z}{q}),
\end{array}
\end{equation}
where $A_{N+1}(z)$, $B(z)$, $F(z)$ are as in (\ref{eq:Vn_NR_AB}) and $G(z)=\sum_{i=0}^{N} g_i z^{i}$, $g_N=c_1$.

\noindent
(b) Time evolution equation
\begin{equation}\label{eq:Vn_Da_T4_ev}
\begin{array}l
\ds G(z)\u{G}(z)=\frac{qc_1c_2A_{N+1}(z)B(z)}{z-a_{N+1}}, \quad {\rm for} \quad F(z)=0,\\[5mm]
\ds F(z)\o{F}(z)=qc_1c_2A_{N+1}(z)B(z), \quad {\rm for} \quad G(z)=0,\\[5mm]
\ds f_0\o{f}_0=\Big(g_0+\frac{d_1A(0)}{a_{N+1}}\Big)\Big(g_0+\frac{d_2A(0)}{a_{N+1}}\Big),
\end{array}
\end{equation}
where $2N$ variables $\frac{f_1}{f_0}$,$\ldots$, $\frac{f_N}{f_0}$, $g_0$,$\ldots$, $g_{N-1}$ are the dependent variables. 

\noindent
(c) The $L_1$ equation
\begin{equation}\label{eq:Vn_Da_T4_L1}
\begin{array}{l}
\ds L_1(z)=A(z)F(\frac{z}{q})y(q z)+qc_1c_2B(\frac{z}{q})F(z)y(\frac{z}{q})\\[5mm]
\ds \phantom{L_1(z)} -\Big\{(z-a_{N+1})F(\frac{z}{q})G(z)+\dfrac{F(z)}{G(\frac{z}{q})}V(\frac{z}{q})\Big\}y(z),
\end{array}
\end{equation}
where $V(z)=qc_1c_2A_{N+1}(z)B(z)-F(z)\o{F}(z)$. 

\subsubsection{Relations among variables $g_i$ and $h$ for each direction
}\label{subsubsec:Vn_Da_rel}

\begin{thm}\label{thm:Vn_NR_Gtrans}
The $L_1$ equations for each direction $T_1$, $\ldots$, $T_4$:
(\ref{eq:Vn_Da_T1_L1}), (\ref{eq:Vn_Da_T2_L1}),  (\ref{eq:Vn_Da_T3_L1}) and  (\ref{eq:Vn_Da_T4_L1})
are equivalent with each other
if the coefficients $g_i$ in $G(z)$ and $h$  are related as
\begin{equation}\label{eq:Vn_NR_Gtrans}
\begin{array}l
\ds \frac{(z-b_1)G(z)^{\S\ref{subsubsec:Vn_Da_T1}}}{A_1(z)}
= \frac{(z-b_1)(z-b_2)}{A_{1,2}(z)}\Big(\frac{G(z)}{1+hz}\Big)^{\S\ref{subsubsec:Vn_Da_T2}}
\\[5mm]
\phantom{
\ds \frac{(z-b_1)G(z)^{\S\ref{subsubsec:Vn_Da_T1}}}{A_1(z)}
}
\ds
=\frac{G(z)^{\S\ref{subsubsec:Vn_Da_T3}}}{A(z)}=\frac{G(z)^{\S\ref{subsubsec:Vn_Da_T4}}}{A_{N+1}(z)},
 \quad \mbox{for $F(z)=0$}.
\end{array}
\end{equation}
\sq
\end{thm}

\prf
The relation (\ref{eq:Vn_NR_Gtrans}) is obtained by
comparing ratios $\frac{y(qz)}{y(z)}$ in $L_1$ under the condition $F(z)=0$.
Then the equivalence of the $L_1$ equations can be checked using their characteristic properties
 by a similar argument as the Appendix \ref{subsec:Lax_Corres}.  \qed
\section{Reduction to $q$-$E_7^{(1)}$ system}\label{sec:Rn}

In this section we consider some $q$-Painlev\'e equations of type $E_7^{(1)}$ as a reduction of $q$-Garnier systems.
\subsection{Notation}\label{subsec:Rn_NR}
Throughout this section, we consider the case $N=3$ and specialize parameters as
\begin{equation}\label{eq:Rn_NR_cons}
c_1=c_2, \quad d_1=d_2.
\end{equation}
Among the directions $T_i$ considered in previous section, the directions consistent with this specialization are the following three:
\begin{equation}\label{eq:Rn_T}
T_1=T_{a_1}^{-1}T_{b_1}^{-1}, \quad T_2=T_{a_1}^{-1}T_{a_2}^{-1}T_{b_1}^{-1}T_{b_2}^{-1}, 
\end{equation}
and inconsistent ones $T_3$ and $T_4$ will be omitted. Under the specialization (\ref{eq:Rn_NR_cons}), we can and will impose constraints on dependent variables as
\begin{equation}\label{eq:Rn_NR_rn}
\begin{array}l
f_0=f_3=0, \quad f_1=w, \quad f_2=-fw,
\end{array}
\end{equation}
where $f$ is one of the unknown variables for the reduced system and $w$ is a gauge freedom. In the followings, we also impose additional constraints (\ref{eq:Rn_Da_T1_rn}) or (\ref{eq:Rn_Da_T2_rn}) in order to reduce the variables $g_i$ and $h$ of $q$-Garnier systems to a variable $g$ and a parameter $e$.
Note that the meaning of the parameter $e$ and the reduced variable $g$ are different depending on directions. The results of \S\ref{subsubsec:Rn_Da_T1} are known in \cite{NY16, Nagao17-2} and those of \S\ref{subsubsec:Rn_Da_T2} are new.
\subsection{Results for each direction}\label{subsec:Rn_Da}

\subsubsection{Reduction from $q$-Garnier system in \S\ref{subsubsec:Vn_Da_T1}}\label{subsubsec:Rn_Da_T1}
\noindent
The direction is  $T_1=T_{a_1}^{-1}T_{b_1}^{-1}$. In this case
we can and will impose an additional condition:
\begin{equation}\label{eq:Rn_Da_T1_rn}
\begin{array}l
g_0=e, \quad g_1=eg, \quad  g_2=c_1,
\end{array}
\end{equation}
where $e=d_1a_2a_3a_4b_1^{-1}$. 

\noindent
(a) Scalar Lax pair

Under the conditions (\ref{eq:Rn_NR_cons}), (\ref{eq:Rn_NR_rn}) and (\ref{eq:Rn_Da_T1_rn}), 
we can reduce the Lax pair (\ref{eq:Vn_Da_T1_L2L3}) to the following Lax pair:
\begin{equation}\label{eq:Rn_Da_T1_L2L3}
\begin{array}l
\ds L_2(z)=wz(1-fz)\o{y}(z)-\prod_{i=2}^{4}(z-a_i)y(qz)+(z-b_1)(e+egz+c_1z^2)y(z),
\\[5mm]
\ds L_3(z)=\o{w}\frac{z}{q}(1-\o{f}\frac{z}{q})y(z)+(z-a_1)\{e+eg\frac{z}{q}+c_1(\frac{z}{q})^2\}\o{y}(z)-qc_1^2\prod_{i=2}^4(\frac{z}{q}-b_i)\o{y}(\frac{z}{q}).
\end{array}
\end{equation}
The Lax pair (\ref{eq:Rn_Da_T1_L2L3}) has been given in \cite{Nagao17-2}.

\noindent
(b) Time evolution equation and equation for $w$
\begin{equation}\label{eq:Rn_Da_T1_ev}
\begin{array}l
\ds (\frac{e}{c_1}f^2+\frac{e}{c_1}gf+1)(\frac{e}{qc_1}f^2+\frac{e}{qc_1}\u{g}f+1)=\frac{\prod_{i=2}^{4}(1-a_if)(1-b_if)}{(1-a_1f)(1-b_1f)},\\[5mm]
\ds \frac{z_1^2(1-fz_1)(1-\o{f}z_1)}{z_2^2(1-fz_2)(1-\o{f}z_2)}=\prod_{i=2}^{4}\frac{(z_1-a_i)(z_1-b_i)}{(z_2-a_i)(z_2-b_i)},
\end{array}
\end{equation}
and
\begin{equation}\label{eq:Rn_Da_T1_guage}
\begin{array}l
\ds w\o{w}=\frac{\prod_{j=2}^{4}(z_i-a_j)(z_i-b_j)}{z_i^2(1-fz_i)(1-\o{f}z_i)}, \quad (i=1,2),
\end{array}
\end{equation}
where $z=z_1, z_2$ are solutions of the equation $e+egz+c_1z^2=0$. The bi-rational equation (\ref{eq:Rn_Da_T1_ev}) is equivalent to the variation of $q$-Painlev\'e equation of type $E_7^{(1)}$ in \cite{Nagao17-2,NY16}.

We remark that the configuration of 8 singular points\footnote{For the theory of the configuration of 8 singular points, see \cite{KNY17, Sakai01}.} is given by 2 points on a line $g=\infty$ and 6 points on a parabolic curve $ef^2+egf+c_1=0$ as follows:
\begin{equation}
(f,g)=(\frac{1}{a_1}, \infty), (\frac{1}{b_1}, \infty), (\frac{1}{a_i}, -\frac{1}{a_i}-\frac{a_ic_1}{e}), (\frac{1}{b_i}, -\frac{1}{b_i}-\frac{b_ic_1}{e}),\quad (i=2,3,4). 
\end{equation}

\noindent
(c) The $L_1$ equation
\begin{equation}\label{eq:Rn_Da_T1_L1}
\begin{array}{l}
\ds L_1(z)=\prod_{i=1}^4(z-a_i)(1-\frac{fz}{q})y(qz)+q^2c_1^2\prod_{i=1}^4(\frac{z}{q}-b_i)(1-fz)y(\frac{z}{q})\\[5mm]
 \ds \phantom{\ds L_1(z)} -\Big\{(z-a_1)(z-b_1)(1-\frac{fz}{q})\varphi(z)+\dfrac{q(1-fz)}{\varphi(\frac{z}{q})}V(\frac{z}{q})\Big\}y(z),
\end{array}
\end{equation}
where $V(z)=qc_1^2\prod_{i=2}^4(z-a_i)(z-b_i)-w\o{w}z^2(1-fz)(1-\o{f}z)$ and $\varphi(z)=e+egz+c_1z^2$. The $L_1$ equation (\ref{eq:Rn_Da_T1_L1}) has been given in \cite{Nagao17-2}.

\subsubsection{
Reduction from $q$-Garnier system in \S\ref{subsubsec:Vn_Da_T2}}\label{subsubsec:Rn_Da_T2}
\noindent
The direction is $T_2=T_{a_1}^{-1}T_{a_2}^{-1}T_{b_1}^{-1}T_{b_1}^{-1}$.
We impose an additional condition:
\begin{equation}\label{eq:Rn_Da_T2_rn}
\begin{array}l
\ds g_0=e, \quad g_1=-\frac{e}{g}, \quad  h=-\frac{e}{c_1g},
\end{array}
\end{equation}
where
$e=d_1a_3a_4(b_1b_2)^{-1}$. 

\noindent
(a) Scalar Lax pair

Under the conditions (\ref{eq:Rn_NR_cons}), (\ref{eq:Rn_NR_rn}) and (\ref{eq:Rn_Da_T2_rn}), 
one can reduce the Lax pair (\ref{eq:Vn_Da_T2_L2L3}) to the following Lax pair:
\begin{equation}\label{eq:Rn_Da_T2_L2L3}
\begin{array}l
\ds L_2(z)=wz(1-fz)\o{y}(z)-\prod_{i=3}^4(z-a_i)(1-\frac{ez}{c_1g})y(qz)+e\prod_{i=1}^2(z-b_i)(1-\frac{z}{g})y(z),
\\[5mm]
\ds L_3(z)=\o{w}\frac{z}{q}(1-\o{f}\frac{z}{q})y(z)+e\prod_{i=1}^2(z-a_i)(1-\frac{z}{qg})\o{y}(z)-qc_1^2\prod_{i=3}^4(\frac{z}{q}-b_i)(1-\frac{ez}{c_1g})\o{y}(\frac{z}{q}).
\end{array}
\end{equation}
The scalar Lax pair (\ref{eq:Rn_Da_T2_L2L3}) is equivalent to the one in \cite{KNY17, Yamada11}.

\noindent
(b) Time evolution equation and equation for $w$
\begin{equation}\label{eq:Rn_Da_T2_ev}
\begin{array}l
\ds \frac{(\frac{c_1}{e}fg-1)(\frac{qc_1}{e}f\u{g}-1)}{(fg-1)(f\u{g}-1)}=\frac{\prod_{i=1}^{2}(1-a_if)(1-b_if)}{\prod_{i=3}^4(1-a_if)(1-b_if)},\\[5mm]
\ds \frac{(fg-\frac{e}{c_1})(\o{f}g-\frac{qe}{c_1})}{(fg-1)(\o{f}g-1)}=\frac{\prod_{i=1}^{2}(g-a_i\frac{e}{c_1})(g-b_i\frac{e}{c_1})}{\prod_{i=3}^4(g-a_i)(g-b_i)},
\end{array}
\end{equation}
and
\begin{equation}\label{eq:Rn_Da_T2_guage}
\begin{array}l
\ds w\o{w}=\frac{\prod_{i=3}^4(g-a_i)(g-b_i)}{qg^2(c_1-e)(c_1-qe)(fg-1)(\o{f}g-1)}.
\end{array}
\end{equation}
The bi-rational equation (\ref{eq:Rn_Da_T2_ev}) is equivalent to the $q$-Painlev\'e equation of type $E_7^{(1)}$ in \cite{GR99, KNY17, Sakai01, Yamada11}, and the configuration of 8 singular points is the well-known standard one.

\noindent
(c) The $L_1$ equation
\begin{equation}\label{eq:Rn_Da_T2_L1}
\begin{array}{l}
\ds L_1(z)=\prod_{i=1}^4(z-a_i)(1-\frac{fz}{q})y(qz)+q^2c_1^2\prod_{i=1}^4(\frac{z}{q}-b_i)(1-fz)y(\frac{z}{q})\\[5mm]
\ds \phantom{L_1(z)} -\frac{1}{1+\frac{ez}{c_1g}}\Big\{e\prod_{i=1}^2(z-a_i)(z-b_i)(1-\frac{fz}{q})(1-\frac{z}{g})+\dfrac{q(1-fz)}{e(1-\frac{z}{qg})}V(\frac{z}{q})\Big\}y(z),
\end{array}
\end{equation}
where $V(z)=qc_1^2(1-\frac{ez}{c_1g})(1-\frac{qez}{c_1g})\prod_{i=3}^4(z-a_i)(z-b_i)-w\o{w}z^2(1-fz)(1-\o{f}z)$. 

\begin{rem}
The $L_1$ equations for each directions $T_1$ and $T_2$ are
equivalent  under the relation \cite{Nagao17-2}
\begin{equation}\label{eq:Rn_NR_gtrans}
\begin{array}l
\ds 
\frac{(1-a_2f)(1-b_2f)}{(\frac{e}{c_1}f^2+\frac{e}{c_1}gf+1)^{\S\ref{subsubsec:Rn_Da_T1}}}\Big(\frac{1-fg}{1-\frac{c_1}{e} fg}\Big)^{\S\ref{subsubsec:Rn_Da_T2}}=1.
\end{array}
\end{equation}
This result follows as a reduction of Theorem \ref{thm:Vn_NR_Gtrans}.
A direct proof is given in \cite{Nagao17-2}.
\sq
\end{rem}

\begin{rem}
If we include more directions such as 
$T_{a_1}^{-1}T_{a_2}^{-1}T_{c_1}^{-1}$ and $T_{a_1}^{-1}T_{a_2}^{-1}T_{d_1}$ in addition to $T_1$ as fundamental ones, 
then all the other directions can be obtained by compositions of them. Thanks to the symmetry of the parameters, the equations for such additional fundamental directions are similar to that of $T_1$. Therefore the results of this section are enough to obtain all the directions of $q$-Painlev\'e equation of type $E_7^{(1)}$.\sq
\end{rem}
\section{Relation to Suzuki's system}\label{sec:KP}

\subsection{$q$-KP and its duality}\label{subsec:SD_Dual}
We first recall a Lax formalism of a certain isomonodromic system formulated as a reduction of the $q$-KP hierarchy \cite{KNY02-2}\footnote{For the related works on $q$-KP, see \cite{KNY02-1} (affine Weyl group symmetry), \cite{Masuda03, Takenawa03} (the case $(m,n)=(2,3), (2,4)$), \cite{KOTY03, Yamada01} (Yang-Baxter maps), \cite{Tsuda10} ($q$-UC hierarchy), \cite{Doliwa14, Hasegawa13} (non-commutative version).} . 

Let $x_{i,j}, r_i, t_j \in \C$ and $m, n \in \N$ be parameters. For an unknown $n$-vector $Y(z)$, we 
consider a linear $q$-difference equation of $n \times n$ matrix system
\begin{equation}\label{eq:SD_mat_L}
Y(qz)={\mathcal A}(z)Y(z), 
\end{equation}
and its deformation system
\begin{equation}\label{eq:SD_mat_D}
\o{Y}(z)={\mathcal B}(z)Y(z).
\end{equation}
Here\footnote{For convenience, extra parameters $r_i$ and $t_j$ are introduced though there are redundancy.}
\begin{equation}\label{eq:SD_mat_AX}
\begin{array}l
{\mathcal A}(z)=dX_m \cdots X_2X_1,\\[3mm]
X_i(z)= \begin{bmatrix}x_{i,1}&1\\&x_{i,2}&1\\&&\ddots&\ddots\\&&&x_{i,n-1}&1\\r_i^{-1} z&&&&x_{i,n}\end{bmatrix}, \quad d={\rm diag}[t_1,\ldots ,t_n].
\end{array}
\end{equation}
In this section, we focus consideration on the linear equation (\ref{eq:SD_mat_L}).

Extending the result of \cite{KNY02-1, KNY02-2}, one obtains the following:
\begin{thm}\label{thm:SD_Dual}
Two cases $(m,n)$ and $(n,m)$ of the linear equation (\ref{eq:SD_mat_L}) are equivalent under the following transformation: 
\begin{equation}\label{eq:SD_mat_Dual_trans}
x_{i,j} \leftrightarrow -x_{j,i},  \quad r_k \leftrightarrow t_k, \quad z \leftrightarrow T_z. 
\end{equation}
\sq
\end{thm}

\prf
Setting $Y_1=Y, Y_{i+1}=X_iY_i$ $(1\leq i \leq m)$ and defining components of $Y$ by $Y_{i,j}=(Y_i)_j$, then we have relations
\begin{equation}\label{eq:SD_mat_Dual_Y}
\begin{array}l
Y_{i+1,j}=x_{i,j} Y_{i,j}+Y_{i,j+1}, \quad Y_{m+1,j}=t_j^{-1}T_z Y_{1,j}, \quad Y_{i,n+1}=r_i^{-1}z Y_{i,1}.
\end{array}
\end{equation}
Applying a transformation (\ref{eq:SD_mat_Dual_trans}) and
\begin{equation}\label{eq:SD_mat_Dual_trans2}
m \leftrightarrow n, \quad Y_{i,j} \leftrightarrow Y_{j,i},
\end{equation}
into the relations (\ref{eq:SD_mat_Dual_Y}), then the cases $(m,n)$ and $(n,m)$ of the linear equation  (\ref{eq:SD_mat_L}) can be transformed with each other.
\qed

We remark that the transformation $z \leftrightarrow T_z$ in (\ref{eq:SD_mat_Dual_trans}) is a kind of $q$-Laplace transformation, and it transforms $T_z z=qzT_z$ into $zT_z=qT_z z$ i.e. $T_z^{-1}z=qzT_z^{-1}$.

\subsection{$q$-Garnier system as the case $(m,n)=(2N+2,2)$ }\label{subsec:SD_Gar}
 
Consider the case $(m,n)=(2N+2,2)$ of the matrix ${\mathcal A}$ given by
\begin{equation}\label{eq:SD_Gar_AX}
\begin{array}l
{\mathcal A}(z)=dX_{2N+2}\cdots X_2X_1,\\[5mm] 
X_i=\left[\begin{array}{cc} x_{i,1}&1\\ r_i^{-1}z&x_{i,2}\end{array}\right], \quad d={\rm diag}[t_1, t_2].
\end{array}
\end{equation}

\begin{prop}\label{prop:SD_Gar}
The matrix ${\mathcal A}$  (\ref{eq:SD_Gar_AX}) is equivalent to the Sakai's matrix ${\mathcal A}$ (\ref{eq:Gar_mat_A}).
\end{prop}
\prf
The matrix ${\mathcal A}$  (\ref{eq:SD_Gar_AX}) takes the form
\begin{equation}\label{eq:SD_Gar_A}
\begin{array}l
{\rm (i)} \quad {\mathcal A}(z)=\sum_{i=0}^{N+1}A_i z^i, \\[5mm]
 {\rm (ii)} \quad A_{N+1}=\left[\begin{array}{cc}t_1\prod_{i=1}^{N+1}r_{2i-1}^{-1}&0\\ *&t_2\prod_{i=1}^{N+1}r_{2i}^{-1}\end{array}\right], \quad
 A_0=\left[\begin{array}{cc}\theta_1&*\\ 0&\theta_2\end{array}\right], \\[5mm]
{\rm(iii)} \quad {\rm det}{\mathcal A}(z)=t_1t_2\prod_{i=1}^{2N+2}r_i^{-1}(z-\alpha_i),
\end{array}
\end{equation}
where
$\theta_j=t_j \prod_{i=1}^{2N+2}x_{i,j}$,
$\alpha_i=r_ix_{i,1}x_{i,2}$.
These properties (\ref{eq:SD_Gar_A}) corresponds exactly to the properties (\ref{eq:Gar_mat_A}) for Sakai's matrix ${\mathcal A}$,
by a gauge transformation with lower triangular matrix $\left[\begin{array}{cc} 1&0\\ *&1\end{array}\right]$. 

Conversely, any matrix ${\mathcal A}$ with properties (\ref{eq:SD_Gar_A}) can be factorized into the form ${\mathcal A}$ (\ref{eq:SD_Gar_AX}). Note that the variables $x_{i,j}$ can be obtained by considering the Ker${\mathcal A}$ at $z=\alpha_i$ due to the property (iii). 
\qed

\begin{rem}
In the case of differential Garnier system, its relation to the KP hierarchy is not so clear.
However, it is known that the Garnier system can been derived through a similarity reduction of the UC hierarchy \cite{Tsuda14}. \sq
\end{rem}

\subsection{Suzuki's system as the case $(m,n)=(2,2N+2)$}\label{subsec:SD_HOP}

Consider the case $(m,n)=(2,2N+2)$ of the matrix ${\mathcal A}$ (\ref{eq:SD_mat_AX}):
\begin{equation}\label{eq:SD_HOP_AX}
\begin{array}l
{\mathcal A}(z)=dX_2X_1, \\[5mm]
X_i = \begin{bmatrix}x_{i,1}&1\\&x_{i,2}&1\\&&\ddots&\ddots\\&&&x_{i,2N+1}&1\\r_i ^{-1}z&&&&x_{i,2N+2}\end{bmatrix}, \quad d={\rm diag}[t_1,\ldots ,t_{2N+2}].
\end{array}
\end{equation}

\begin{prop}\label{prop:SD_HOP}
The matrix ${\mathcal A}$ (\ref{eq:SD_HOP_AX}) has the following form:
\begin{equation}\label{eq:SD_HOP_A}
\begin{array}l
{\rm(i)}\quad {\mathcal A}(z)= \begin{bmatrix}\alpha_1&\varphi_1&t_1\\&\alpha_2&\varphi_2&t_2\\&&\ddots&\ddots&\ddots\\&&&\alpha_{2N}&\varphi_{2N}&t_{2N}\\ \frac{t_{2N+1}}{r_1}z&&&&\alpha_{2N+1}&\varphi_{2N+1}\\ \varphi_{2N+2}z&\frac{t_{2N+2}}{r_2}z&&&&\alpha_{2N+2}\end{bmatrix},\\[20mm]
{\rm(ii)}\quad {\rm det}{\mathcal A}(z)=r_1^{-1}r_2^{-1}\prod_{i=1}^{2N+2}t_i(z-\theta_1)(z-\theta_2).
\end{array}
\end{equation}
Conversely, any matrix ${\mathcal A}$ of the form (\ref{eq:SD_HOP_A}) can be factorized as (\ref{eq:SD_HOP_AX}).
\sq
\end{prop}
\prf
It is easy to check that the matrix ${\mathcal A}$ (\ref{eq:SD_HOP_AX}) takes the form (\ref{eq:SD_HOP_A})
under substitutions 
$\alpha_i=t_ix_{1,i}x_{2,i}$,  $\theta_i=r_i\prod_{j=1}^{2N+2}x_{i,j}$,
$\varphi_{k}=t_k(x_{1,1+k}+x_{2,k}) \quad (k\neq 2N+2)$ and
$\varphi_{2N+2}=t_{2N+2}(r_2^{-1}x_{1,1}+r_1^{-1}x_{2,2N+2})$.

Conversely, the matrix ${\mathcal A}$ (\ref{eq:SD_HOP_A}) can be factorized as ${\mathcal A}$ (\ref{eq:SD_HOP_AX}); consider the Ker ${\mathcal A}$ at $z=\theta_i$ using the property (ii). 
\qed

\begin{rem}\label{rem:SD_HOP_Suzuki}
In \cite{Suzuki17} T.Suzuki considered the matrix ${\mathcal A}$ (\ref{eq:SD_HOP_A}), with redundant parameters fixed as $r_1=\frac{1}{t}$, $r_2=1$, $t_1=\ldots=t_{2N+2}=1$. 
Note that two variables among $\varphi_1$, $\dots$, $\varphi_{2N+2}$ can be reduced by a gauge transformation and the constraint coming from the property (ii). 
He also gave the  matrix ${\mathcal B}$ corresponding with a time evolution $\o{r_1}=qr_1$ (i.e. $\o{t}=\frac{t}{q}$), $\o{\theta_1}=q\theta_1$. \sq
\end{rem}
 
Thanks to Theorem \ref{thm:SD_Dual} and Proposition \ref{prop:SD_Gar}, Proposition \ref{prop:SD_HOP}, we have the conclusion that
the $q$-Garnier system and Suzuki's higher order Painlev\'e system are related with each other in the sense that they are the deformations of the linear equations equivalent under a $q$-Laplace transformation.

By the correspondence of parameters $(n=N+1)$
\begin{equation}\label{eq:SD_HOP_corres}
\begin{array}{c|cccccc|c}
\mbox{\rm Matrix}&& {\rm Parameters}&&&&&{\rm Specialization}\\
\hline
(\ref{eq:Gar_mat_A}) &a_1, \ldots, a_{n}&b_1,\ldots,b_{n}&c_1&c_2&d_1&d_2&\\
(\ref{eq:SD_Gar_AX}) &\alpha_1, \ldots, \alpha_{n}&\alpha_{n+1},\ldots,\alpha_{2n}&t_1&\frac{t_2}{q}&\frac{\theta_1}{\prod_{i=1}^{n}(-\alpha_i)}&\frac{\theta_2}{\prod_{i=1}^{n}(-\alpha_i)}&r_1=\ldots=r_{2n}=1\\
(\ref{eq:SD_HOP_A}) &\alpha_1,\ldots,\alpha_{n}&\alpha_{n+1},\ldots,\alpha_{2n}&r_1&\frac{r_2}{q}&\frac{\theta_1}{\prod_{i=1}^{n}(-\alpha_i)}&\frac{\theta_2}{\prod_{i=1}^{n}(-\alpha_i)}&t_1=\ldots=t_{2n}=1,
\end{array}
\end{equation}
the Suzuki's direction $\o{r}_1=qr_1$, $\o{\theta}_1=q\theta_1$ for (\ref{eq:SD_HOP_A}) corresponds to $T_3=T_{c_1}T_{d_1}$ in \S\ref{subsubsec:Vn_Da_T3}. 

\section{Summary}\label{sec:sum}
In this paper, extending the previous work \cite{NY16}, we obtained the following two main results.

\begin{itemize}
\item Choosing the proper variables according to several deformation directions, we obtained simple expression of the scalar Lax pairs and the evolution equations for the variations of the $q$-Garnier system. Consequently, we gave relations among these $q$-Garnier systems. As a byproduct, we obtained the standard $q$-Painlev\'e equations of type $E_7^{(1)}$ as a reduction of a certain $q$-Garnier system.


\item We clarified the relation between the $q$-Garnier system \cite{NY16, Sakai05-1} and the Suzuki's system \cite{Suzuki17} by formulating the duality of the $q$-KP system.
\end{itemize}

\appendix
\section{Lax equations}\label{sec:Lax}
\subsection{Correspondence to Sakai's Lax form}\label{subsec:Lax_Corres}
The $L_1$ equations in (\ref{eq:Gar_Sca_L}) and those in \S\ref{subsec:Vn_Da} are all equivalent with each other. In this appendix, we give the correspondence\footnote{Other cases are similar. The case of (\ref{eq:Gar_Sca_L}) and (\ref{eq:Vn_Da_T1_L1}) is given in \cite[\S2.2]{NY16}.} in case of  (\ref{eq:Gar_Sca_L}) and (\ref{eq:Vn_Da_T3_L1}) as an example.

We call $q^{\rho_0}$ (resp. $q^{\rho_{\infty}}$) characteristic exponents at $x=0$ (resp. $x=\infty$) when  solutions $y(x)$ have the form 
\begin{equation}\label{eq:y0}
y(x)=k_{\rho_0}x^{\rho_0}(1+O(x)), \quad \mbox{at $x=0$},
\end{equation}
\begin{equation}\label{eq:yinf}
y(x)=k_{\rho_{\infty}}x^{\rho_{\infty}}(1+O\Big(\frac{1}{x}\Big)), \quad \mbox{at $x=\infty$}.
\end{equation}
Eliminating $\o{y}(z)$ and $\o{y}(\frac{z}{q})$ from $L_2(z)= L_2(\frac{z}{q})=L_3(z)=0$ (\ref{eq:Vn_Da_T3_L2L3}), we obtain  $L_1(z)=0$ (\ref{eq:Vn_Da_T3_L1}). Then we have

\begin{lem}\label{lem:Lax_Corres_T3L1}
The $L_1$ equation (\ref{eq:Vn_Da_T3_L1}) takes the following properties: 

(i) It is a linear three term equation between $y(qz)$, $y(z)$ and $y(\frac{z}{q})$, and the coefficients of $y(qz)$, $y(z)$ and $y(\frac{z}{q})$ are polynomials of degree $2N+1$ in $z$,

(ii) The coefficient of $y(qz)$ (resp. $y(\frac{z}{q})$) has zeros at $z=a_1, \cdots , a_{N+1}$ (resp. $z=qb_1, \cdots , qb_{N+1}$),

(iii) The characteristic exponents of solutions $y(z)$ are $d_1, d_2$ (at $z=0$) and $c_1, c_2$ (at $z=\infty$) generically\footnote{Generic case means $f_0f_N\ne 0$. We remark that the reduced cases to $q$-$E_7^{(1)}$ system are nongeneric, where their $L_1$ equations (\ref{eq:Rn_Da_T1_L1}) and (\ref{eq:Rn_Da_T2_L1}) have exponents $d_1$, $qd_1$ (at $z=0$) and $c_1$, $\frac{c_1}{q}$ ($z=\infty$).}. 

(iv) The $N$ points $z$ such that $F(z)=0$ are apparent singularities (i.e., the solutions are regular there),  where 
\begin{equation}\label{eq:Vn_T4_evef1}
\dfrac{y(qz)}{y(z)}=\dfrac{G(z)}{A(z)}, \quad {\mbox for}\quad F(z)=0,
\end{equation}
holds (c.f. (\ref{eq:Gar_Sca_apparent})).

Conversely, the $L_1$ equation (\ref{eq:Vn_Da_T3_L1}) is uniquely characterized by these properties (i)$-$(iv).

\sq
\end{lem}

On the other hand, owing to Lemma \ref{lem:Gar_Sca_L}, Sakai's Lax equation (\ref{eq:Gar_mat_L}) is rewritten into the $L_1$ equation (\ref{eq:Gar_Sca_L}). Furthermore we have
\begin{lem}\label{lem:Lax_Corres_gauge}
The $L_1$ equation (\ref{eq:Gar_Sca_L}) can be expressed as a form
\begin{equation}\label{eq:Lax_Corres_L1}
\begin{array}l
\ds L_1: \prod_{i=1}^{N+1}(z-\alpha_i)F(\frac{z}{q})y(qz)-\left\{F(\frac{z}{q}){\mathcal A_{11}}(z)+F(z){\mathcal A_{22}}(z)\right\}y(z)
\\
\phantom{\ds L_1: \prod_{i=1}^{N+1}(z-\alpha_i)F(\frac{z}{q})y(qz)}
\ds +\kappa_1\kappa_2\prod_{i=N+2}^{2N+2}(\frac{x}{q}-\alpha_i)F(z)y(\frac{z}{q})=0.
\end{array}
\end{equation}
\sq
\end{lem}
\prf
The proof is given by using a gauge transformation: 
\begin{equation}\label{eq:Lax_Corres_gauge}
\ds y(z) \mapsto H(z)y(z), \quad \mbox{with} \quad \frac{H(qz)}{H(z)}=\prod_{i=1}^{N+1}(z-\alpha_i).
\end{equation} 
\qed

Similarly to Lemma \ref{lem:Lax_Corres_T3L1}, we have
\begin{lem}\label{lem:Lax_Corres_Sakai}
Then the $L_1$ equation (\ref{eq:Lax_Corres_L1}) has the following properties:

(i) It is a linear three term equation between $y(qz)$, $y(z)$ and $y_1(\frac{z}{q})$, and the coefficients of $y(qz)$, $y(z)$ and $y(\frac{z}{q})$ are polynomials of degree $2N+1$ in $z$, 

(ii) The coefficient of $y(qz)$ (resp. $y(\frac{z}{q})$) has zeros at $z=\alpha_1, \cdots , \alpha_{N+1}$ (resp. $z=q \alpha_{N+2}, \cdots , q \alpha_{2N+2}$). 

(iii) The characteristic exponents of the solutions $y(z)$ are $\frac{\theta_1}{\prod_{i=1}^{N+1}(-\alpha_i)}, \frac{\theta_2}{\prod_{i=1}^{N+1}(-\alpha_i)}$ (at $z=0$) and $\kappa_1, q^{-1}\kappa_2$ (at $z=\infty$), 

(iv) $N$ points $z=\lambda_i$ such that $F(z)=0$ are apparent singularities, where 
\begin{equation}\label{eq:Lax_Corres_A11def}
\frac{y(qz)}{y(z)}=\dfrac{A_{11}(z)}{\prod_{i=1}^{N+1}(z-\alpha_i)}, \quad {\mbox for}\quad F(z)=0,
\end{equation}
holds. 

Conversely, the $L_1$ equation (\ref{eq:Lax_Corres_L1}) is uniquely characterized by these properties (i)$-$(iv). 

\sq
\end{lem}

Hence we obtain
\begin{prop}
The $L_1$ equations (\ref{eq:Vn_Da_T3_L1}) and (\ref{eq:Gar_Sca_L}) are equivalent up to the gauge transformation (\ref{eq:Lax_Corres_gauge}) and changes of variables and parameters.
 \sq
\end{prop}
\prf
Thanks to Lemmas \ref{lem:Lax_Corres_T3L1}, \ref{lem:Lax_Corres_gauge} and \ref{lem:Lax_Corres_Sakai}.
\qed
 
\subsection{Sufficiency for the compatibility}\label{subsec:Lax_Suf}
In \S\ref{subsec:Vn_Da}, we stated that the equations in item (b) are sufficient for the compatibility of the equations in item (a). Here we will prove this  
fact in the case of \S\ref{subsubsec:Vn_Da_T3} as an example\footnote{Other cases can be proved in the similar way and the case of \S\ref{subsubsec:Vn_Da_T1} has been done in \cite[\S2.3]{NY16}.}.

Eliminating ${y}(z)$ and  ${y}(qz)$ from $L_2(z)=L_3(z)=L_3(qz)=0$ (\ref{eq:Vn_Da_T3_L2L3}), we obtain the following expression:
\begin{equation}\label{eq:Vn_T3_L4}
\begin{array}{l}
\ds L_4(z)=\o{A}(z)\o{F}(\frac{z}{q})\o{y}(q z)+q^2c_1c_2\o{B}(\frac{z}{q})\o{F}(z)\o{y}(\frac{z}{q})\\[5mm]
\ds -\Big\{q\o{F}(z)G(\frac{z}{q})+\dfrac{\o{F}(\frac{z}{q})}{G(z)}V(z)\Big\}\o{y}(x),
\end{array}
\end{equation}
where $V$ is given in (\ref{eq:Vn_Da_T3_L1}).

\begin{lem}\label{lem:Lax_Suf_T4L4}
The $L_4$ equation (\ref{eq:Vn_T3_L4}) has the following properties: 

(i) It is a linear three term equation between $\o{y}(qz)$, $\o{y}(z)$ and $\o{y}(\frac{z}{q})$,and the coefficients of $\o{y}(qz)$, $\o{y}(z)$ and $\o{y}(\frac{z}{q})$ are polynomials of degree $2N+1$ in $z$,

(ii) The coefficient of $\o{y}(qz)$ (resp. $\o{y}(\frac{z}{q})$) has zeros at $z=a_1, a_2 , \cdots , a_{N+1}$ (resp. $z=qb_1, qb_2 , \cdots , qb_{N+1}$),

(iii) The characteristic exponents of solutions $\o{y}(z)$ are $qd_1, d_2$ (at $z=0$) and $qc_1, c_2$ (at $z=\infty$),

(iv) The $N$ points $z$ such that $\o{F}(z)=0$ are apparent singularities, where 
\begin{equation}\label{eq:Vn_T4_evef2}
\dfrac{\o{y}(qz)}{\o{y}(z)}=\dfrac{qc_1c_2B(z)}{G(z)}, \quad {\mbox for}\quad \o{F}(z)=0,
\end{equation}
holds. 

Conversely, the equation $L_4=0$ (\ref{eq:Vn_T3_L4}) is uniquely characterized by these properties (i)$-$(iv).
\sq
\end{lem}

\begin{prop}\label{prop:Lax_Suf_com}
The linear $q$-difference equations $L_1$ (\ref{eq:Vn_Da_T3_L1}) and $L_2$ (\ref{eq:Vn_Da_T3_L2L3}) are compatible
if and only if the bi-rational equation (\ref{eq:Vn_Da_T3_ev}).
\sq
\end{prop}

\prf
The compatibility of $L_1$ and $L_2$ means that $T(L_1)=L_4$. This can be checked by the characterizations of the equations $L_1$ (resp. $L_4$) in Lemma \ref{lem:Lax_Corres_T3L1} (resp. \ref{lem:Lax_Suf_T4L4}). \qed

\section{From Pad\'e interpolation to the $q$-Garnier system }\label{sec:Pade}

There is a convenient method to approach the continuous/discrete Painlev\'e equations 
by using certain problem of Pad\'e approximation \cite{Nagao17-1,Yamada09-1} (see  also \cite{Mano12, MT17}) or  interpolation 
\cite{Ikawa13,Nagao15,Nagao16, Nagao17-2,Nagao17-3,NTY13,Yamada14,Yamada17}. 
In \cite{NY16}, this method is applied to the $q$-Garnier system in the case of  \S\ref{subsubsec:Vn_Da_T1}.
Here, we illustrate the derivation of the Lax pair  (\ref{eq:Vn_Da_T3_L2L3}) through the Pad\'e method in case of $q$-Garnier system in \S\ref{subsubsec:Vn_Da_T3} as an example. 

\subsection{Derivation of scalar Lax pair}\label{subsec:Pade_Lax}
Fix a positive integer $N \in \N$ and a parameter $q$ ($0 < |q| <1$). For parameters $a_1, \ldots, a_{N}, b_1, \ldots, b_{N}$ and $c$, we consider a function
\begin{equation}\label{eq:Pade_Lax_psi}
\psi(z)=c^{\log_q z}\prod_{i=1}^{N}\dfrac{(a_i z, b_i)_\infty}{(a_i, b_i z)_\infty}.
\end{equation}
Let $P(z)$ and $Q(z)$ be polynomials of degree $m$ and $n$ $\in \Z_{\ge 0}$ in $z$ determined by the following Pad\'e interpolation condition:
\begin{equation}\label{eq:Pade_Lax_pade}
\psi(z_s)=\dfrac{P(z_s)}{Q(z_s)}, \quad (z_s=q^s, s=0, 1, \ldots m+n).
\end{equation}
The common normalizations of the polynomials $P$ and $Q$ in $z$ are fixed as $P(0)=1$. We consider the shift operation: 
$\o{x}=T(x)$ where $T=T_c$.

We construct  two linear $q$-difference relations: $L_2=0$ among $y(z), y(qz), \o{y}(z)$ and $L_3=0$ among $y(z), \o{y}(z), \o{y}(\frac{z}{q})$ satisfied by the functions $y=P$ and $y=\psi Q$.
\begin{prop}
The linear relations $L_2$ and $L_3$ can be expressed as
\begin{equation}\label{eq:Pade_Lax_L2L3} 
\begin{array}l
\ds L_2(z)=F(z)\o{y}(z)+A(z)(\frac{z}{q^{m+n}})_1y(qz)+\frac{c}{g_0}G(z)y(z)=0,\\[3mm]
\ds L_3(z)=\frac{z}{q}\o{F}(\frac{z}{q})y(x)+\frac{1}{g_0}G(\frac{z}{q})\o{y}(z)-(z)_1 B(\frac{z}{q})\o{y}(\frac{z}{q})=0,
\end{array}
\end{equation}
where $A(z)=\prod_{i=1}^{N}(a_i z)_1$, $B(z)=\prod_{i=1}^{N}(b_i z)_1$ and $F(z)$, $G(z)$ are as in (\ref{eq:Vn_Da_T3_L2L3}) and $f_0, \ldots ,f_N$, $g_0, \ldots , g_N$ are constants depending on parameters $a_i, b_j, c$ and $m, n \in \Z_{\ge 0}$. 

\sq
\end{prop}

\prf
By the definition above, the linear relations $L_2$ and $L_3$ can be written as
{\small
\begin{equation}\label{eq:Pade_Lax_mat}
\begin{array}l
L_2(z)\propto
\begin{vmatrix}
y(z) & y(qz) & \o{y}(z) \\
{\bf y}(z) & {\bf y}(qz) & \o{\bf y}(z)
\end{vmatrix}=D_1(z) \o{y}(z)-D_2(z)y(qz)+D_3(z)y(z)=0, 
\\ L_3(z)\propto
\begin{vmatrix} 
y(z) & \o{y}(z) & \o{y}(\frac{z}{q}) \\
{\bf y}(z) & \o{\bf y}(z) & \o{\bf y}(\frac{z}{q})
\end{vmatrix}=\o{D}_1(\frac{z}{q})y(z)+D_3(\frac{z}{q}) \o{y}(z)-D_2(z)\o{y}(\frac{z}{q})=0,
\end{array}
\end{equation}
}
where ${\bf y}(z)=\left[\begin{array}{c}P(z)\\ \psi(z)Q(z)\end{array}\right]$ and Casorati determinants 
\begin{equation}\label{eq:Pade_Lax_D}
\begin{array}{l}
D_1(z)=|{\bf y}(z),{\bf y}(qz)|, \quad  D_2(z)=|{\bf y}(z),{\o{\bf y}}(z)|, \quad D_3(z)=|{\bf y}(qz),\o{{\bf y}}(z)|.
\end{array}
\end{equation}
Using the relations
\begin{equation}\label{eq:Pade_Lax_ratio}
\dfrac{\psi(qz)}{\psi(z)}=c\dfrac{B(z)}{A(z)},
\quad \dfrac{\o{\psi}(z)}{\psi(z)}=z,
\end{equation}
we can rewrite the Casorati determinants (\ref{eq:Pade_Lax_D}) into the following determinants:
\begin{equation}\label{eq:Pade_Lax_DD}
\begin{array}{l}
\ds D_1(z)=\dfrac{\psi(z)}{A(z)}\left\{cB(z)P(z)Q(qz)-A(z)P(qz)Q(z)\right\}
=:\dfrac{\psi(z)}{A(z)}\prod_{i=0}^{m+n-1}(\frac{z}{q^i})_1c_0F(z),\\
\ds D_2(z)=\psi(z)\left\{zP(z)\o{Q}(z)-\o{P}(z)Q(z)\right\}
=:\psi(z)\prod_{i=0}^{m+n}(\frac{z}{q^i})_1c_0,\\
\ds D_3(z)=\dfrac{\psi(z)}{A(z)}\left\{zA(z)P(qz)\o{Q}(z)-cB(z)\o{P}(z)Q(qz)\right\}
=:\dfrac{\psi(z)}{A(z)}\prod_{i=0}^{m+n-1}(\frac{z}{q^i})_1G(z).
\end{array}
\end{equation}
Then the constants $c_0=\dfrac{g_0}{c}$, 
$g_{N+1}=-\dfrac{g_0\prod_{i=1}^N(-a_i)}{cq^n}$
are determined through the expansions around $x=0$ and $x=\infty$. As a result, we obtain the desired equations. \qed

\vskip10mm
\noindent
{\bf Acknowledgment.} 
The authors shall thank Professors Kenji Kajiwara, Tetsu Masuda, Masatoshi Noumi, Hidetaka Sakai, Takao Suzuki and Teruhisa Tsuda for stimulating comments. 
This work is partially supported by JSPS KAKENHI (26287018) and Expenses Revitalizing Education and Research of Akashi College (0217030).


\vskip5mm

\begin{thebibliography}{A}
\bibitem{Doliwa14}
Doliwa A., {\it Non-cmmutative rational Yang-Baxter maps}, Lett. Math. Phys., {\bf 104}, (2014), 299--309.
\bibitem{DST13}
Dzhamay A., Sakai H., and Takenawa T., {\it Discrete Hamiltonian Structure of Schlesinger Transformations}, arXiv:1302.2972 [math-ph].
\bibitem{DT14}
Dzhamay A., and Takenawa T., {\it Geometric Analysis of Reductions from Schlesinger transformations to difference Painlev\'e equations}, arXiv:1408.3778 [math-ph].
\bibitem{GaR04}
Gasper G., and Rahman M., {\it Basic Hypergeometric Series. With a foreword by Richard Askey. Second edition. Encyclopedia of Mathematics and its Applications}, {\bf 96}. Cambridge University Press Cambridge, (2004). 
\bibitem{GR99}
Grammaticos B., Ramani A., On a novel $q$-discrete analogue of the Painlev\'{e} {\rm VI} equation, Phys. Lett. A, {\bf257} (1999), 288--292. 
\bibitem{Hasegawa13}
Hasegawa K., {\it Quantizing the discrete Painlev\'e {\rm VI} equation: the Lax formalism}, Lett. Math. Phys., {\bf 103} (2013), 865--879.
\bibitem{Ikawa13}
Ikawa Y., {\it Hypergeometric Solutions for the $q$-Painlev\'e Equation of Type $E_6^{(1)}$ by the Pad\'e method}, Lett. Math. Phys., {\bf 103} (2013), 743--763.
\bibitem{JS96}
Jimbo M. and Sakai H., {\it A $q$-analog of the sixth Painlev\'e equation}, Lett. Math. Phys., {\bf 38} (1996), 145--154.
\bibitem{KNY02-1}
Kajiwara K., Noumi M., and Yamada Y.,
{\it Discrete dynamical systems with W($A_{m-1}^{(1)} \times A_{n-1}^{(1)}$) symmetry}, Lett. Math. Phys., {\bf 60} (2002),  211--219.
\bibitem{KNY02-2}
Kajiwara K., Noumi M., and Yamada Y., 
{\it $q$-Painlev\'e Systems Arising from $q$-KP Hierarchy}, Lett. Math. Phys., {\bf 62} (2002), 259--268.
\bibitem{KNY17}
Kajiwara K., Noumi M., and Yamada Y.,
{\it Geometric aspects of Painlev\'e equations}, J. Phys. A: Math. Theor., {\bf 50} (2017), 073001(164pp) (Topical Review). 
\bibitem{KOTY03}
Kuniba A., Okado M., Takagi T., and Yamada Y.,  {\it Geometric crystal and tropical R for $D_{n}^{(1)}$}, Int. Math. Res. Not., {\bf 5} (2003), 2565--2620.
\bibitem{Mano12}
Mano T., {\it Determinant formula for solutions of the Garnier system and Pad\'e approximation}.
J. Phys. A: Math. Theor., {\bf 45} (2012), 135206--135219.
\bibitem{MT17}
Mano T., and Tsuda T., {\it Hermite-Pad\'e approximation, isomonodromic deformation and hypergeometric integral}. Math. Z., {\bf 285} (2017), no. 1-2, 397--431.
\bibitem{Masuda03}
Masuda T., {\it On the rational solutions of $q$-Painlev\'e {\rm V} equation}, Nagoya Math. J., {\bf 169} (2003), 119--143.
\bibitem{Masuda09}
Masuda T., {\it Hypergeometric $\tau$-functions of the $q$-Painlev\'e system of type $E^{(1)}_7$}, SIGMA, {\bf 5} (2009), 035 (30pp).
\bibitem{Nagao15}
Nagao H., {\it The Pad\'e interpolation method applied to $q$-Painlev\'e equations}. Lett. Math. Phys., {\bf 105} (2015),  503--521.
\bibitem{Nagao16}
Nagao H., {\it  Lax pairs for additive difference Painlev\'e equations}, arXiv:1604.02530 [nlin.SI].
\bibitem{Nagao17-1}
Nagao H., {\it The Pad\'e interpolation method applied to $q$-Painlev\'e equations II (differential grid version)}, Lett. Math. Phys., {\bf 107} (2017), 107--127.
\bibitem{Nagao17-2}
Nagao H., {\it A variation of the $q$-Painlev\'e system with affine Weyl group symmetry of type $E_7^{(1)}$}, arXiv:1706.10087 [nlin.SI].
\bibitem{Nagao17-3}
Nagao H., {\it Hypergeometric special solutions for $d$-Painlev\'e equations}, arXiv:1706.10101 [nlin.SI].
\bibitem{NY16}
Nagao H., and Yamada Y., {\it Study of $q$-Garnier system by Pad\'e method}, Funkcialaj Ekvacioj, to appear, arXiv:1601.01099 [nlin.SI].
\bibitem{ND16}
Nijhoff F., and Delice N., {\it On elliptic Lax pairs and isomonodromic deformation systems for elliptic lattice equations}, arXiv:1605.00829 [nlin.SI].
\bibitem{NTY13}
Noumi M., Tsujimoto S., and Yamada Y., {\it Pad\'{e} Interpolation for Elliptic Painlev\'e Equation}, Springer {\bf 40} (2013), 463--482.
\bibitem{OR16-1}
Ormerod, C.M., and Rains E.M., {\it Commutation Relations and Discrete Garnier Systems}, SIGMA, {\bf 12} (2016), 110, 50 pages. 
\bibitem{OR16-2}
Ormerod, C.M., and Rains E.M., {\it An elliptic Garnier system}, arXiv:1607.07831 [nlin.SI]. 
\bibitem{Sakai01}
Sakai H., {\it Rational surfaces with affine root systems and geometry of the Painlev\'e equations}, Commun. Math. Phys. {\bf 220} (2001), 165--221.
\bibitem{Sakai05-1} Sakai H., {\it A $q$-analog of the Garnier system}, Funkcialaj Ekvacioj, {\bf 48} (2005), 273--297.
\bibitem{Sakai05-2} Sakai H., {\it Hypergeometric Solution of $q$-Schlesinger System of Rank Two}, Lett. Math. Phys., 
{\bf 73} (2005), 237--247.
\bibitem{Sakai06}
Sakai H., {\it Lax form of the $q$-Painlev\'e equation associated with the $A_2^{(1)}$ surface}, J. Phys. A: Math. Gen., {\bf 39} (2006), 12203--12210.
\bibitem{Suzuki15}
Suzuki T., {\it A $q$-analogue of the Drinfeld-Sokolov hierarchy of type $A$ and $q$-Painlev\'e system}, AMS Contemp. Math., {\bf 651} (2015), 25--38.
\bibitem{Suzuki17}
Suzuki T., {\it A reformulation of the generalized $q$-Painlev\'e {\rm VI} system with $W(A_{2n+1}^{(1)})$ symmetry}, J. Integrable Syst., {\bf 2} (2017), 1--18.
\bibitem{Takenawa03}
Takenawa T., {\it Weyl group symmetry of type $D_5^{(1)}$ in the $q$-Painlev\'e {\rm V} equation}, Funkcialaj Ekvacoj, {\bf 46} (2003), 173--186.
\bibitem{Tsuda10}
Tsuda T., {\it On an integrable system of q-difference equations satisfied by the universal characters: its Lax formalism and an application to $q$-Painlev\'e equations}, Comm. Math. Phys., {\bf 293} (2010), 347--359.
\bibitem{Tsuda14}
Tsuda T., {\it UC hierarchy and monodromy preserving deformation}, J. reine angew. Math., {\bf 690} (2014), 1--34.
\bibitem{Yamada01}
Yamada Y., {\it A birational representation of Weyl group, combinatorial R-matrix and discrete Toda equation}, in "Physics and combinatorics, 2000", World Sci. Publ., (2001), 305--319.
\bibitem{Yamada09-1}
Yamada Y., {\it Pad\'e method to Painlev\'e equations}, Funkcialaj Ekvacoj, {\bf 52} (2009), 83--92.
\bibitem{Yamada11}
Yamada Y., {\it Lax formalism for $q$-Painlev\'e equations with affine Weyl group symmetry of type $E^{(1)}_n$}, Int. Math. Res. Not., {\bf 17} (2011), 3823--3838.
\bibitem{Yamada14}
Yamada Y., {\it A simple expression for discrete Painlev\'e equations}, RIMS Kokyuroku Bessatsu, {\bf B47} (2014), 087--095.
\bibitem{Yamada17}
Yamada Y., {\it An elliptic Garnier system from interpolation}, arXiv:1706.05155 [math-ph].

\end{thebibliography}
\end{document}